# Linear Network Coding on Multi-Mesh of Trees (MMT) using All to All Broadcast (AAB)

Nitin Rakesh[1] and VipinTyagi[2]

[1, 2] Department of CSE & IT, Jaypee University of Information Technology, Solan, H.P. 173215, India

**Abstract**
We introduce linear network coding on parallel architecture for multi-source finite acyclic network. In this problem, different messages in diverse time periods are broadcast and every non-source node in the network decodes and encodes the message based on further communication.We wish to minimize the communication steps and time complexity involved in transfer of data from node-to-node during parallel communication.We have used Multi-Mesh of Trees (MMT) topology for implementing network coding. To envisage our result, we use all-to-all broadcast as communication algorithm.

## 1. Introduction

Shuo-Yen *et al*. [1] prove constructively that by linear coding alone, the rate at which a message reaches each node can achieve the individual max-flow bound. Also, provide realization of transmission scheme and practically construct linear coding approaches for both cyclic and acyclic networks. [1–6] shows that network coding is necessity to multicast two bits per unit time from a source to destinations. It also showed that the output flow at a given node is obtained as a linear combination of its input flows. The content of any information flowing out of a set of non-source nodes can be derived from the accumulated information that has flown into the set of nodes. Shuo-Yen *et al*. described an approach on network information flow and improved the performance in data broadcasting in all-to-all communication in order to increase the capacity or the throughput of the network.

[7] selected the linear coefficients in a finite field of opportune size in a random way. In this paper packetizing and buffering are explained in terms of encoding vector and buffering the received packets. It showed that each node sends packets obtained as a random linear combination of packets stored in its buffer and each node receives packets which are linear combinations of source packets and it stores them into a matrix. While Widmer *et. al*. [8] gave an approach with energy efficient broadcasting in network coding. Subsequent work by Fragouli *et. al*. [9] gave two heuristics and stated that each node in the graph is associated with a forwarding factor. A source node transmits its source symbols (or packets) with some parameters bounded by this forwarding factor. And when a node receives an innovative symbol, it broadcast a linear combination over the span of the received coding vector. [10] deals with network coding of a single message over an acyclic network. Network coding over undirected networks was introduced by [11] and this work was followed by [12], [13]. The network codes that involve only linear mapping with a combination of global and local encoding mapping involves linear error-correction code [14], [15], [16] and [17] have also been presented.

We present an approach for parallel network in which network coding is employed to perform communication amid the nodes. The association among network coding and communication algorithm establishes a more efficient way to transfer data among the nodes. We consider parallel multi-source multicast framework with correlated sources on MMT architecture [18]. We use a randomized approach in which, other than the receiving nodes all nodes perform random linear mapping from inputs on outputs (see figure 1). In each incoming transmissions from source node, the destination node has knowledge of overall linear combination of that data set from source. This information is updated at each coding node, by applying the same linear mapping to the coefficient vectors as applied to the information signals. As an example, assume that in a directed parallel network ($Ň$) the source node $P_1$ (unique node, without any incoming at that instant of time) sends a set of two bits ($d_1$, $d_2$) to node $P_2$, $P_3$ and $P_4$, $P_7$ (figure 1).

Network ($Ň$) in figure 1 is used to show information multicast with network coding at each node. Node $P_1$ multicast data set ($d_1$, $d_2$) to destination nodes $P_3$ and $P_7$. Any receiving non-destination node, truncheons randomly chosen coefficient of finite fields with the received data before transferring to other nodes. The compressive transmission throughout network ($Ň$) is heralded in following steps of Table 1.

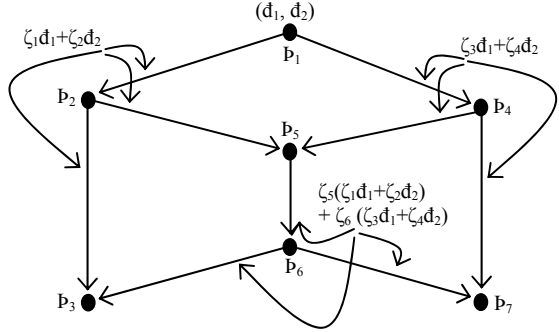

Fig. 1. A network ($\check{N}$) used, as an example, to explain LNC with coefficient added at each data transfer from different nodes (the network has seven nodes $P_1, P_2...P_7$ and nine edges $P_1 P_2, P_1 P_4, P_2 P_5, P_4 P_5, P_2 P_3, P_5 P_6, P_4 P_5, P_6 P_3, P_6 P_7$ directed in this order). ($d_1, d_2$) is the set of data being multicast to destinations, and coefficients $\zeta_1, \zeta_2...\zeta_6$ are randomly chosen elements of a finite field. Each link represents the data transmission.

Table 1: Compressive transmission in Network ($\check{N}$).

| S. No | Source Node | Destination Node | Coefficient Clubbed | Data to send further |
|---|---|---|---|---|
| 1 | $P_1$ | $P_2$ and $P_4$ | ($\zeta_1, \zeta_2$ and $\zeta_3, \zeta_4$) | $\zeta_1 d_1+\zeta_2 d_2$ and $\zeta_3 d_1+\zeta_4 d_2$ |
| 2 | $P_2, P_4$ | $P_5$ | ($\zeta_5$ and $\zeta_6$) | $\zeta_5(\zeta_1 d_1+\zeta_2 d_2)+\zeta_6(\zeta_3 d_1+\zeta_4 d_2)$ |
| 3 | $P_5$ | $P_6$ | | $\zeta_5(\zeta_1 d_1+\zeta_2 d_2)+\zeta_6(\zeta_3 d_1+\zeta_4 d_2)$ |
| 4 | $P_6$ | $P_7$ | | ($d_1, d_2$) |

This approach indicates that using network coding efficient multicast of diverse data in a network is possible. It is also right to say that the flow of information and flow of physical commodities are two different things [1]. So, the coding of information does not increase the information content. The capacity of a network to transmit information from the source to destination can become higher if the coding scheme becomes wider but it is limited to max-flow for a very wide coding scheme [1].

Now, as parallel networks contribute in data communication within several nodes in parallel, it is required to have higher capacity of such networks to transmit information. In senseto increase capacity of data communication in parallel networks we have implemented network coding on parallel architecture (MMT). To examine the performance of this network with network coding we have implemented this approach with existing All-to-All Broadcast algorithm (AAB) [19] on this architecture. In consecutive sections we have also shown that this approach has reduced the chance of error and increased the capacity of network to transmit data between nodes. For parallel transmission of information linearity of coding makes encoding (at source) and decoding (at receiving end) easy to contrivance. We do not address the problem which may or may not occur because of this approach, but have identified the possibilities of errors after implementation.

The remaining paper is organized in five sections. In section second our basic model and preliminaries are illustrated. In basic model, notions used for *linear–code multicast in parallel architecture* (LCM-PA) and definitions are explained. In section third, implementation of AAB on parallel network-MMT is explained. In section forth, LNC is implemented using LCM-PA, it is illustrated using AAB algorithm on MMT [19]. The fifth section is used for results and simulations. In section sixth, we are concluding this paper and future scope of LCM-PA is given in this section. The basic definition, theorems and lemma used in this paper are explained in appendix.

## 2. Model and Preliminaries

A parallel network is represented as a directed graph° G (V, E), where V is the set of nodes in network and E is the set of links, such that, from node $i$ to $j$ for all $(i, j) \in E$, where node $i$ and $j$ are called the *origin* and *destination*, respectively, of link $(i, j)$, information can be sent noiselessly. Each link $l \in E$ is associated with a nonnegative real number $c_l$ representing its transmission capacity in bits per unit time. The origin and destination of a link $l \in E$ are denoted as $o(l)$ and $d(l)$, respectively, where $o(l) \neq d(l) \forall l \in E$ is obtained as a coding function of information received at $o(l)$.

There are $r$ discrete memoryless information source processes $X_1, X_2...X_r$, which are random binary sequences. The processors may change according to the parallel architecture. We denote the Slepian–Wolf region of the sources
$\mathcal{R}_{SW} = \{(R_1, R_2, ..., R_r) : \sum_{i \in S} R_i > H(X_S | X_{S^c}) \forall S \subseteq \{1, 2, ..., r\}\}$
Where, $X_S = (X_{i_1}, X_{i_2}, ..., X_{i_{|S|}})$, $i_k \in S, k = 1... |S|$. Source process $X_i$ is generated at node $a(i)$, and multi-cast to all nodes $j \in b(i)$, where $a : \{1,..., r\} \to V$ and $b : \{1,..., r\} \to 2^V$ are arbitrary mappings. For parallel architectures, we have considered the same approach to implement the network coding. In this paper, we consider the (multisource) multicast case where $b(i) = \{\beta_1,..., \beta_d\}$ for all $i \in [1, r]$. The node $a(1),..., a(r)$ are called *source nodes* and the $\beta_1,..., \beta_d$ are called receiver nodes, or receivers. For simplicity, we assume subsequently that $a(i) \neq \beta_j \forall i \in [1, r], j \in [1, d]$. For data communication at different step, the source and destination nodes changes according to the flow of data in the algorithm. If the receiving node is able to encode the complete source information, than connection requirements are fulfilled. For parallel communication, we have used these sets of connection requirements for level of communication, which is encoded at each level (step of algorithm). To specify a multicast connection problem we used 1) a graph G (V, E), 2) a set of multicast connection requirements, and 3) a set of link capacity $\{c_l \mid l \in E\}$. To explain *linear–code multicast* (LCM) with parallel communication network, we present some terminologies, definitions and assumptions.

*Conventions:* 1) In MMT network, the edges e.g., $(P_1, P_2) \in (E)$ denotes that $(P_1, P_2)$ is a bi-directed edge [18], but this edge may act as unidirectional depending on the algorithm. 2) The information unit is taken as a symbol in the base field, i.e., 1 symbol in the base field can be transmitted on a channel every unit time [1].

*Definitions:* 1) The communication in MMT network is interblock and intrablock [18]. LNC is implemented in blocks first and then in complete network. 2) A LCM on a communication network $(G, o(l))$ is an assignment of vector space $v(P_i)$ to every node $P_i$ and a vector $v(P_i, P_j)$ to every edge $P_i, P_j$ [1] such that
$v(o(l)) = \Omega$;
$v(P_i, P_j) \in v(P_i)$ for every edge $P_i, P_j$;
for any collection $\in d(l)$ of nonsource nodes in the network
$\langle\{v(P_l): P_l \in d(l)\}\rangle = \langle\{v(P_i, P_j): P_i \notin d(l), P_j \in d(l)\}\rangle$

*Assumptions:* 1) Each source process $X_i$ has one bit per unit time entropy rate for independent source process, while larger rate sources are modeled as multiple sources. 2) Modeling of sources as linear combinations of independent source processes for linearly correlated sources. 3) Links with $l \in E$ is supposed having a capacity $c_l$ of one bit per unit time for both independent as well as linear correlated sources. 4) Both cyclic (networks with link delays because of information buffering at intermediate nodes; operated in a batched [2] fashion, burst [11], or pipelined [12]) and acyclic networks (networks whose nodes are delay-free i.e. zero-delay) are considered for implementation of LNC on parallel networks, by analyzing parallel network to be a cyclic or acyclic. 5) We are repeatedly using either of these terms processor and nodes, throughout the paper, which signify common significance.

The network may be analyzed to be acyclic or cyclic using scalar algebraic network coding framework [13]. Let us consider the zero-delay case first, by representing the equation $Y_j = \sum_{\{i:a(i)=o(j)\}} a_{i,j} X_i + \sum_{\{l:d(l)=o(j)\}} f_{l,j} Y_l$. The sequence of length-$u$ blocks or vectors of bits, which are treated as elements of a finite field $F_q$, $q = 2^u$. The information process $Y_j$ transmission on a link $j$ is formed as a linear combination, in $F_q$, of link $j$'s inputs, i.e., source processes $X_i$ for which $a(i) = o(j)$ and random processes $Y_l$ for which $d(l) = o(j)$. The $i$th output process $Z_{\beta,i}$ at receiver node $\beta$ is a linear combination of the information processes on its terminal links, represented as
$Z_{\beta,i} = \sum_{\{l:d(l)=\beta\}} b_{\beta,i,l} Y_l$.

Memory is needed, for link delays on network for multicast, at receiver (or source) nodes, but a memoryless operation suffices at all other nodes [12]. The linear coding equation for unit delay links (considered) are

$Y_j(t+1) = \sum_{\{i:a(i)=o(j)\}} a_{i,j} X_i(t) + \sum_{\{l:d(l)=o(j)\}} f_{l,j} Y_l(t)$.
$Z_{\beta,i}(t+1) =$
$\sum_{u=0}^{\mu} Å_{\beta,i}(u) Z_{\beta,i}(t-u) + \sum_{\{l:d(l)=\beta\}} \sum_{u=0}^{\mu} Ä_{\beta,i,l}(u) Y_l(t-u)$

where $X_i(t)$, $Y_j(t)$, $Z_{\beta,i}(t)$, $Å_{\beta,i}(t)$, and $Ä_{\beta,i,l}(t)$ are the values of variables at $t$ time and represents the required memory. In terms of delay variable $D$ these equation are as $Y_j(D) = \sum_{\{i:a(i)=o(j)\}} D a_{i,j} X_i(D) + \sum_{\{l:d(l)=o(j)\}} D f_{l,j} Y_l(t)$.

$$Z_{\beta,i}(D) = \sum_{\{l:d(l)=\beta\}} b_{\beta,i,l}(D) Y_l(D).$$

where

$$b_{\beta,i,l}(D) = \frac{\sum_{u=0}^{\mu} D^{u+1} Ä_{\beta,i,l}(u)}{1 - \sum_{u=0}^{\mu} D^{u+1} \sum_{u=0}^{\mu} Å_{\beta,i}(u)}$$

and $X_i(D) = \sum_{t=0}^{\infty} X_i(t) D^t$

$$Y_j(D) = \sum_{t=0}^{\infty} Y_t(j) D^t, \qquad Y_j(0) = 0$$

$$Z_{\beta,i}(D) = \sum_{t=0}^{\infty} Z_{\beta,i}(D)(t) D^t, \quad Z_{\beta,i}(0) = 0$$

The above given coefficients can be collected into $r \times |E|$ matrices. These coefficients can be used from the transmission in parallel network. These matrices will be formed for both cyclic and acyclic cases.

$$A = \begin{cases} (a_{i,j}) \text{ in the acyclic delay-free case} \\ (D a_{i,j}) \text{ in the cyclic case with delay} \end{cases}$$

And $B = (b_{\beta,i,l})$, and the matrix $|E| \times |E|$

$$F = \begin{cases} (f_{l,j}) \text{ in the acyclic delay-free case} \\ (D f_{l,j}) \text{ in the cyclic case with delay} \end{cases}$$

Now, let us consider an example of parallel network ($\check{N}$) (MMT), in which processor $P_1$ (unique processor, without any incoming at that instant of time) to node $P_2$ and $P_3$, sends two bits, $(d_1, d_2)$ as given in figure 2.

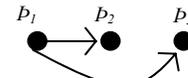

Fig. 2. A row of a block of MMT with $n = 3$, where $n$ is the number of processors in MMT architecture. The detailed MMT architecture is given in figure 3 for more simplicity to the readers.

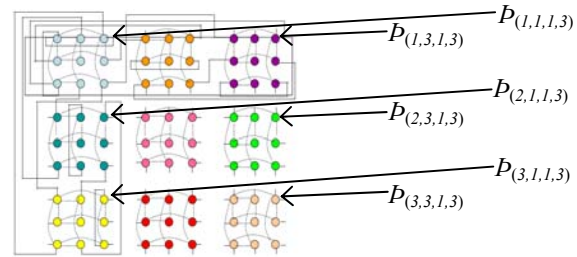

Fig. 3. 3×3 Multi-Mesh of Trees (MMT) ($\check{N}$). (All interblock links are not shown. The $P_{(1,3,1,3)}$, $P_{(2,3,1,3)}$, $P_{(3,3,1,3)}$ are the processor index value which is used to identify individual processors through-out the architecture).

This linear coding is achieved with the LCM-PA ($\psi$, used to replace LCM-PA further for equations) is specified by

$$\psi(P_1, P_2) = \psi(P_1, P_3)\begin{pmatrix}1\\1\end{pmatrix}$$

The matrix product of $(d_1\ d_2)$ with the column vector assigned by $\psi$ is the data sent in a row of MMT. Further, for $n$ number of processors the data received by other processors will be $d_1 + d_2$, where vector $d_1 + d_2$ reduce to the exclusive-OR $d_1 \oplus d_2$. Also, for every $\psi$ on a network, for all nodes $P_3$ (which is the receiving processor) [1]

$$\dim(v(P_3) \leq \text{maxflow}(P_3).$$

This shows that maxflow($P_3$) is an upper bound on the amount of information received at $P_3$ when a LCM $\psi$ is used [1].

## 3. Implementation of AAB on Parallel Network

In this section, we implement AAB on parallel network (MMT). For implementation, we are using AAB algorithm, which involves ten steps to completely transfer and receive information of all processors to all processors in MMT [19] and implement LNC using LCM-PA model in next section. We consider the MMT network with $n = 8$, where $n$ is number of processors and in algorithm and we consider $N = n^2 \times n^2$, $\forall n \in N$ and a *block = n×n = row× column*. The time taken to transfer and receive all information at each step of algorithm is listed in [19] involved in AAB algorithm.

For implementation of AAB on MMT network, first we state a reason for using this network. MMT network is better than other traditional parallel networks (we compared few of them e.g., Multi-Mesh (MM) [20, 21]) based on the topological properties of MMT, which is comparable regarding efficiency parameters. A comparison of these networks, based on some parameters, is given in figure 4 and a comparison between 2D Sort on MM and MMT for different values of processor is given in figure 5.

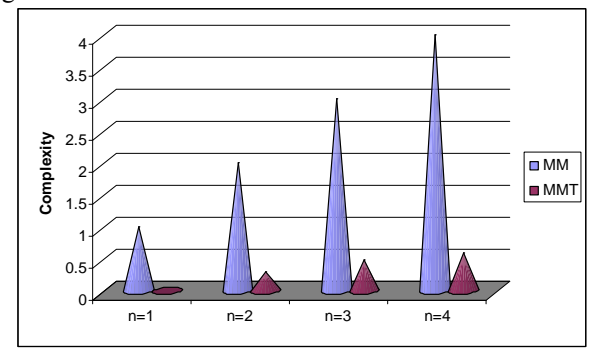

Fig. 4. Comparison of MMT and MM on the basis of Communication links, Solution of Polynomial Equations, One to All and Row & Column Broadcast.

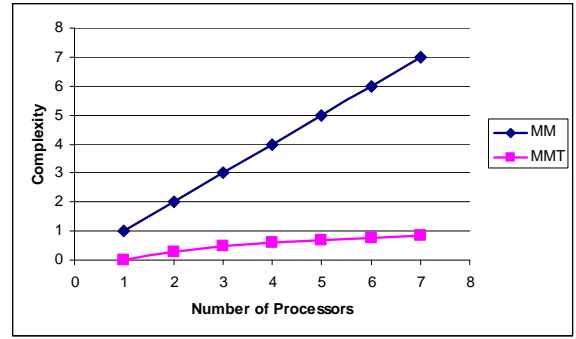

Fig. 5. A comparison between 2D Sort on MM and MMT for different values of processor.

Table 2 [18, 22] shows characteristics of various processor organizations based on some of the network optimization parameters. From all these network architectures MMT is more optimum network to be used.

Table 2: Characteristics of Various Processor Organizations.

| Network | Nodes | Diameter | Bisection Width | Constant Number of Edges | Constant Edge Length |
|---|---|---|---|---|---|
| 1-D mesh | $k$ | $k-1$ | 1 | Yes | Yes |
| 2-D mesh | $k^2$ | $2(k-1)$ | $k$ | Yes | Yes |
| 3-D mesh | $k^3$ | $3(k-1)$ | $k^2$ | Yes | Yes |
| Binary tree | $2^k - 1$ | $2(k-1)$ | 1 | Yes | No |
| 4-ary hypertree | $2^k(2^k - 1)$ | $2k$ | $2^{k+1}$ | Yes | No |
| Pyramid | $4k^2 - 1)/3$ | $2\log k$ | $2k$ | Yes | No |
| Butterfly | $(k + 1)2^k$ | $2k$ | $2^k$ | Yes | No |
| Hypercube | $2^k$ | $k$ | $2^{k-1}$ | No | No |
| Cube-connected cycles | $k2^k$ | $2k$ | $2^{k-1}$ | Yes | No |
| Shuffle-exchange | $2^k$ | $2k-1$ | $\geq 2^{k-1}/k$ | Yes | No |
| De Bruijn | $2^k$ | $k$ | $2^k/k$ | Yes | No |
| MMT | $k^4$ | $4\log k + 2$ | $2(k-1)$ | Yes | No |
| MM | $k^4$ | $2k$ | $2(k-1)$ | No | No |

Now, to demonstrate the algorithm, we consider $N = 8^2 \times 8^2 = 4096$ nodes, as the size of network, where each *block* consists of $8 \times 8$ i.e., *row × column*. For clarity in explaining each step of algorithm, we have used either one row or one column, based on the algorithm, to show the flow of data in each step. For every step the data flow varies, so for each step different algorithms are used. Figure 6 shows first row in first block of the network and the connectivity between the processors is based on the topological properties of MMT [18]. We have considered that each processor is having a Working Array (*WA*) which consist of the processor index ($P_n$) and information associated with that processor ($I_n$). The size of working array is based on the size of network used, i.e. for $n = 8$, the size of $WA = 8$.

466

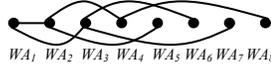

Fig. 6. Shows initial condition of processors containing WA(only one row of a block of 8 × 8 MMT is shown)

The figure 7 (a) shows the position of data after completion of step 1 and figure 7 (b) shows the content of $WA_1$ after step 1.

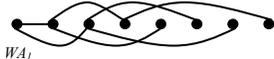

| P_1 | P_2 | P_2 | P_4 | P_5 | P_6 | P_7 | P_8 |
|---|---|---|---|---|---|---|---|
| I_1 | I_2 | I_3 | I_4 | I_5 | I_6 | I_7 | I_8 |

Fig. 7. (a) After Step 1     (b) Content of $WA_1$ after Step 1

### Algorithm 1. Step 1 of AAB

a. /* This operation is common between all processors of each row of each block,
b. Each node is represented by $n_{(\alpha,\beta,i,j)}$; where $\alpha, \beta$ are the block index and $i, j$ are node index (see figure M)
c. The transfer is conducted in order $\frac{n_{(\alpha,\beta,i,j)}}{2\times Count\ of\ Iteration-1} < j \leq \frac{n_{(\alpha,\beta,i,j)}}{2\times Count\ of\ Iteration-1}$ */

1: Starting from each row of each block of network, such that the processor with greater index value will transfer data to lower index processors linked according to the topological properties of network.
2: repeat
3: Select nodes $n_{(\alpha,\beta,i,\frac{N}{2}+1)}, n_{(\alpha,\beta,i,\frac{N}{2}+2)}, n_{(\alpha,\beta,i,\frac{N}{2}+3)}, \ldots n_{(\alpha,\beta,i,N)} \in N$ from each block of network such that at each transfer the block is divided in two parts (e.g. if $N = 40$, number of nodes in blocks will also be 40 and division will be 1 to 20 and 21 to 40th index position) and transfer message to remaining nodes $n_{(\alpha,\beta,i,1)}, n_{(\alpha,\beta,i,2)}, n_{(\alpha,\beta,i,3)}, \ldots n_{(\alpha,\beta,i,\frac{N}{2})} \in N$ linked according to topological properties of this network.
*Note: The message will be transferred from higher processor index to lower.*
4: Select nodes $n_{(\alpha,\beta,i,1)}, n_{(\alpha,\beta,i,2)}, n_{(\alpha,\beta,i,3)}, \ldots n_{(\alpha,\beta,i,\frac{N}{2})} \in N$ (other than the nodes from which message has already transferred) from each block of network such that at each transfer these nodes are divided in two parts (same as in 3; i.e. $n_{(\alpha,\beta,i,1)}, n_{(\alpha,\beta,i,2)}, \ldots n_{(\alpha,\beta,i,\frac{N}{4})}$, and $n_{(\alpha,\beta,i,\frac{N}{4}+1)}, n_{(\alpha,\beta,i,\frac{N}{4}+2)} \ldots n_{(\alpha,\beta,i,\frac{N}{2})} \in N$. Now $n_{(\alpha,\beta,i,\frac{N}{4}+1)}, n_{(\alpha,\beta,i,\frac{N}{4}+2)} \ldots n_{(\alpha,\beta,i,\frac{N}{2})}$ will transfer respective messages to $n_{(\alpha,\beta,i,1)}, n_{(\alpha,\beta,i,2)}, \ldots n_{(\alpha,\beta,i,\frac{N}{4})}$, linked according to topological properties of this network.
5: until all nodes have finished transmitting and forwarding.

### Algorithm 2. Step 2 of AAB

a. /* This operation is common between all root processors of each row of each block,
b. Root processors of each row of a block are identified as in figure B,
c. The transfer of information of all root processors of respective rows is conducted according to connectivity. */

1: Starting from each row of each block. The root nodes of respective rows will transfer data to connected nodes of that row.
2: repeat
3: until all nodes have received the information of root processors.

After the completion of step 2 the position of data in a row is shown in figure 8. The data from the root node of a row of all blocks of network receives the complete information of that row as the content of $WA_1$.

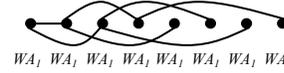

Fig. 8. After Step 2

### Algorithm 3. Step 3 of AAB

a. /* This operation is common between all root processors of each column of each block,
b. The transfer is conducted in order $\frac{n_{(\alpha,\beta,i,j)}}{2\times Count\ of\ Iteration-1} < i \leq \frac{n_{(\alpha,\beta,i,j)}}{2\times Count\ of\ Iteration-1}$ */

1: Starting from each column of each block of network, such that the processor with greater index value will transfer data to lower index processors linked according to the topological properties of network.
2: repeat
3: Select nodes $n_{(\alpha,\beta,i,\frac{N}{2}+1)}, n_{(\alpha,\beta,i,\frac{N}{2}+2)}, n_{(\alpha,\beta,i,\frac{N}{2}+3)}, \ldots n_{(\alpha,\beta,i,N)} \in N$ from each block of network such that at each transfer the block is divided in two parts (e.g. if $N = 40$, number of nodes in blocks will also be 40 and division will be 1 to 20 and 21 to 40th index position) and transfer message to remaining nodes $n_{(\alpha,\beta,i,1)}, n_{(\alpha,\beta,i,2)}, n_{(\alpha,\beta,i,3)}, \ldots n_{(\alpha,\beta,i,\frac{N}{2})} \in N$ linked according to topological properties of this network.

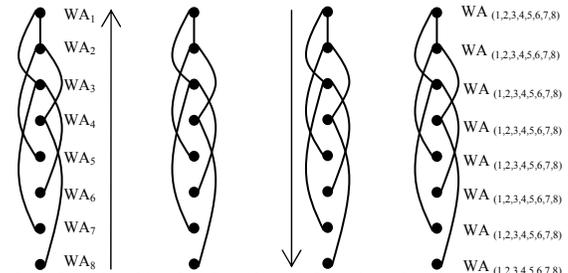

Fig. 9. a) Step 3   b) After Step 3   c) Step 4   d) After Step 4

Figure 9 shows the Step 3 and 4 in which the communication is performed in each column of each block of the network. After the completion of step 4 each column of each block of network consists of complete information of respective column.

### Algorithm 4. Step 4 of AAB

a. /* This operation is common between all root processors of each column of each block,
b. Root processors of each column of each block are identified as in figure C,
c. The transfer of information of all root processors of respective columns is conducted according to connectivity. */

1: Starting from each column of each block. The root nodes of respective columns will transfer data to connected nodes of that row.
2: repeat
3: until all nodes have received the information of root processors.

Algorithm 5. Step 5 of AAB (Interblock Communication)

/* The step is performed using the °horizontal interblock links of this network which transfers the information of all the blocks of respective rows to the root processors of respective block with processor index ($j = N$) */

1: Starting from each blocks of each rows the information is communicated to the root processors of respective block in such a manner that the processor index $n_{(\alpha,\beta,i,j=N)}$.
2: In one communication step this information is broadcasted to every root processor of respective block of respective row. This step is performed on entire network.
*Note: At the end of this step every root processor contains the information of complete block from which this information is broadcasted.*

Algorithm 6. Step 6 of AAB (Interblock Communication)

a. /* This step uses algorithm 3 for communicating the information received after step 5 (algorithm 5).
b. This operation is common between all root processors of each column of each block,
c. The transfer is conducted in order $\frac{n_{(\alpha,\beta,i,j)}}{2 \times Count\ of\ Iteration - 1} < i \leq \frac{n_{(\alpha,\beta,i,j)}}{2 \times Count\ of\ Iteration - 1}$ */

1: Starting from each column of each block of network, such that the processor with greater index value will transfer data to lower index processors linked according to the topological properties of network.
2: repeat
3: Select nodes $n_{(\alpha,\beta,i,\frac{N}{2}+1)}, n_{(\alpha,\beta,i,\frac{N}{2}+2)}, n_{(\alpha,\beta,i,\frac{N}{2}+3)}, \ldots n_{(\alpha,\beta,i,N)} \in N$ from each block of network such that at each transfer the block is divided in two parts and transfer message to remaining nodes $n_{(\alpha,\beta,i,1)}, n_{(\alpha,\beta,i,2)}, n_{(\alpha,\beta,i,3)}, \ldots n_{(\alpha,\beta,i,\frac{N}{2})} \in N$ linked according to topological properties of this network.

Algorithm 7. Step 7 of AAB (Interblock Communication)

/* °One-to-all broadcast is used in the block*/

To transfer the information of a block in a row to other block of respective rows the one-to-all broadcast algorithm is used.
*Note: At the end of this step, complete blocks of each row have information of all processors in that row.*

Algorithm 8. Step 8 of AAB (Interblock Communication)

/* The step is performed using the horizontal interblock links of this network which transfers the information of all the blocks of respective columns to the root processors of respective block with processor index ($i = N$) */

1: Starting from each blocks of each columns the information is communicated to the root processors of respective block in such a manner that the processor index $n_{(\alpha,\beta,i=N,j)}$.
2: In one communication step this information is broadcasted to every root processor of respective block of respective column. This step is performed on entire network.
*Note: At the end of this step every root processor contains the information of complete block from which this information is broadcasted.*

Algorithm 9. Step 9 of AAB

a. /* This operation is common between all processors of each row of each block,
b. Each node is represented by $n_{(\alpha,\beta,i,j)}$;
c. The transfer is conducted in order $\frac{n_{(\alpha,\beta,i,j)}}{2 \times Count\ of\ Iteration - 1} < i \leq \frac{n_{(\alpha,\beta,i,j)}}{2 \times Count\ of\ Iteration - 1}$ */

1: Starting from each row of each block of network, such that the processor with greater index value will transfer data to lower index processors linked according to the topological properties of network.
2: repeat
3: Select nodes $n_{(\alpha,\beta,i,\frac{N}{2}+1)}, n_{(\alpha,\beta,i,\frac{N}{2}+2)}, n_{(\alpha,\beta,i,\frac{N}{2}+3)}, \ldots n_{(\alpha,\beta,i,N)} \in N$ from each block of network such that at each transfer the block is divided in two parts and transfer message to remaining nodes $n_{(\alpha,\beta,i,1)}, n_{(\alpha,\beta,i,2)}, n_{(\alpha,\beta,i,3)}, \ldots n_{(\alpha,\beta,i,\frac{N}{2})} \in N$ linked according to topological properties of this network.
*Note: The message will be transferred from higher processor index to lower.*
4: Select nodes $n_{(\alpha,\beta,i,1)}, n_{(\alpha,\beta,i,2)}, n_{(\alpha,\beta,i,3)}, \ldots n_{(\alpha,\beta,i,\frac{N}{2})} \in N$ (other than the nodes from which message has already transferred) from each block of network such that at each transfer these nodes are divided in two parts (same as in 3; i.e. $n_{(\alpha,\beta,i,1)}, n_{(\alpha,\beta,i,2)}, \ldots n_{(\alpha,\beta,i,\frac{N}{4})}$, and $n_{(\alpha,\beta,i,\frac{N}{4}+1)}, n_{(\alpha,\beta,i,\frac{N}{4}+2)} \ldots n_{(\alpha,\beta,i,\frac{N}{2})} \in N$. Now $n_{(\alpha,\beta,i,\frac{N}{4}+1)}, n_{(\alpha,\beta,i,\frac{N}{4}+2)} \ldots n_{(\alpha,\beta,i,\frac{N}{2})}$ will transfer respective messages to $n_{(\alpha,\beta,i,1)}, n_{(\alpha,\beta,i,2)}, \ldots n_{(\alpha,\beta,i,\frac{N}{4})}$, linked according to topological properties of this network.
5: until all nodes have finished transmitting and forwarding.

Algorithm 10. Step 10 of AAB

/* °AAB is used in the block*/

Select block from each column to transfer information of a block in a column to other block of respective columns for this AAB is used.
*Note: At the end of this step, all the processors of each block contains information of all processors of the network.*

## 4. Implementing LNC on AAB using MMT

In this section we implement network coding for each step to make the communication faster and increase the rate of information transmitted from each node. We consider network as delay-free (acyclic) and $o(l) \neq d(l)$. The algorithm results are analyzed later with $n= 8$ processors.
For each step independent and different algorithms are used (see section IV) and linear coding is implemented with each algorithm. According to algorithm 1, data from all processors are transferred with $n = 8$ and *count* = 1 to 2 i.e, $(8/(2 \times 1 - 1)) <j \leq (8/2 \times 1) = 8 <j \leq 4$, which means the processors $P_1$, $P_2$, $P_3$ and $P_4$ will receive data from $P_5$, $P_6$, $P_7$ and $P_8$, shown in figure 10.

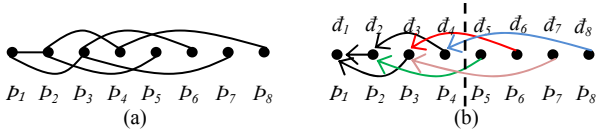

Fig. 10. (a) Shows the indexing of processors with respect to nodes in the figure. (b) Shows the direction of flow of data in step 1 of AAB algorithm on MMT, $P_1$, $P_2$, $P_3$ and $P_4$ are the processor receiving data and $P_5$, $P_6$, $P_7$ and $P_8$ are the sending processors. The dotted line distinguishes between the receiving and sending processors in first iteration of step 1.

*Step 1*: Linear coding is implemented on $P_1$, $P_2$ and $P_3$ processors, as these are receiving a set of data form source processors $P_5$, $P_6$, $P_7$ and $P_8$ in first iteration. Processor $P_8$ is source and $P_4$ is its destination; $P_7$ and $P_6$ are sources and $P_3$ is their destination; lastly in $\log n$ iteration i.e. (3 iteration for $n = 8$), $P_1$ will receive data from $P_2$ and $P_3$. After implementation of LNC according to LCM-PA on these sources and destinations, step 1 will work as in figure 11. During first iteration of AAB on MMT, LCM-PA will work as in figure 11 (a). Data from $P_5$, $P_6$, $P_7$ and $P_8$ is sent to $P_2$, $P_3$, $P_3$ and $P_4$ respectively. So, the complete set of data from all processors reached processor $P_1$, i.e. after execution of step 1 all data, in a row, will reach its root processors, but due to LCM-PA the data reached $P_1$ will have time complexity of $(\log n - 1)$, as one step is reduced during transfer of the data using LCM-PA model.

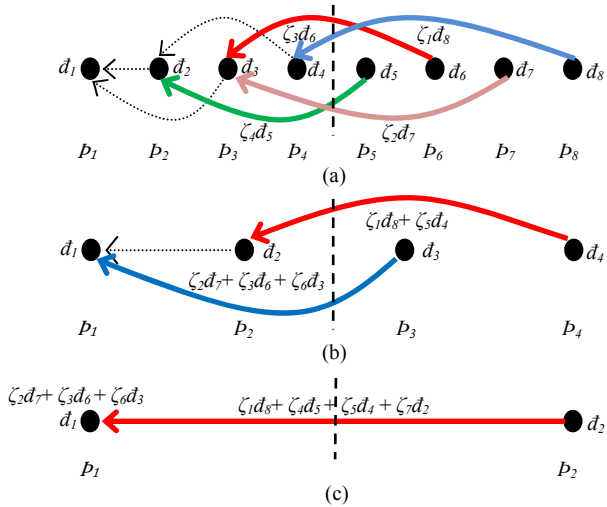

Fig. 11. (a) Iteration first of step 1; data from processors $P_5$, $P_6$, $P_7$ and $P_8$ is sent to processors to $P_4$, $P_3$, $P_3$ and $P_2$ respectively. (b) Iteration second of step 1; data from processors $P_4$ and $P_3$ is sent to processors to $P_2$ and $P_1$ respectively. (c) Iteration third of step 1; data from processors $P_2$ is sent to processors $P_1$.

*Step 2*: The root processors of each row, ($P_1$: root processor of first block and first row) will broadcast the data (from $P_1$: $\zeta_1 d_8 + \zeta_2 d_7 + \zeta_3 d_6 + \zeta_4 d_5 + \zeta_5 d_4 + \zeta_6 d_3 + \zeta_7 d_2$) to all the processors of respective row using intrablock links transfer, see figure 12.

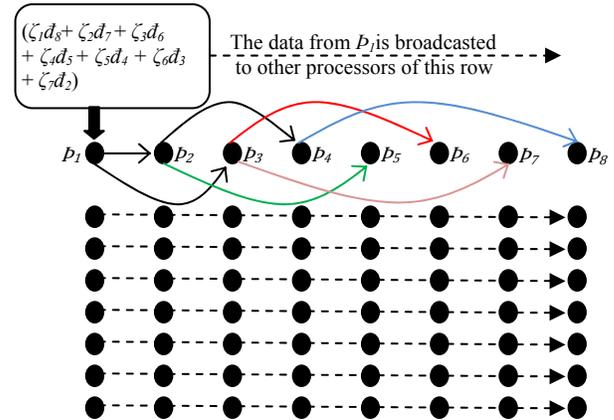

Fig. 12. The data from each row root processor is broadcasted to other processors of respective row in each block.

The time complexity for this step will be reduced by $n$ i.e., $n(\log n - 1)$. This step is a broadcasting step in each block with intrablock links of MMT. At the end of this step, complete data from root processor is received by otherprocessors of that row. LCM-PA is applied at the same level as in step 1, but the size of data increases to $n$.

*Step 3*: This step is similar to step 1, but in this step the data is broadcasted in column-wise order of each block. Linear coding is implemented on $P_{11}$, $P_{12}$ and $P_{13}$ processors, as these are receiving a set of data form source processors $P_{15}$, $P_{16}$, $P_{17}$ and $P_{18}$ in first iteration. Processor $P_{18}$ is source and $P_{14}$ is its destination; $P_{17}$ and $P_{16}$ are sources and $P_{13}$ is their destination; lastly in $\log n$ iteration i.e. (3 iteration for $n = 8$), $P_{11}$ will receive data from $P_{12}$ and $P_{13}$. After implementation LCM-PA on these sources and destinations, step 3 works as in figure 13.

*Step 4*: In this step all the root processors of each column and each block, ($P_{11}$: root processor of first block and first column) will broadcast the data (from $P_{11}$: $\zeta_1 d_{18} + \zeta_2 d_{17} + \zeta_3 d_{16} + \zeta_4 d_{15} + \zeta_5 d_{14} + \zeta_6 d_{13} + \zeta_7 d_{12}$) to all the processors of respective column using intrablock links transfer, see figure 14. The time complexity of this step is reduced by $n^2$ i.e. $n^2(\log n - 1) = n^2 \log n - n^2$. The coefficient value ($\zeta_i$) in step 4 is different from the coefficient value in step 1.

*Step 5*: After step 4, each processors of respective columns contains information of all processors of that column. The step 5, perform the interblock communication using the horizontal interblock links which transfers this information (of all the blocks of respective rows) to the root processors (of respective block), and this requires one communication step (CS) [19]. The time complexity of this step will be same as of AAB i.e. 1CS.

*Step 6:* Using step 3, for transferring information of all the processors in the column at the processors with P_ID ($j=n$), so the WA of all the processors is transferred in the column in the order $n/(2count-1) < j \leq n/(2count)$. Time complexity of step 6: $n^3 \log n$.

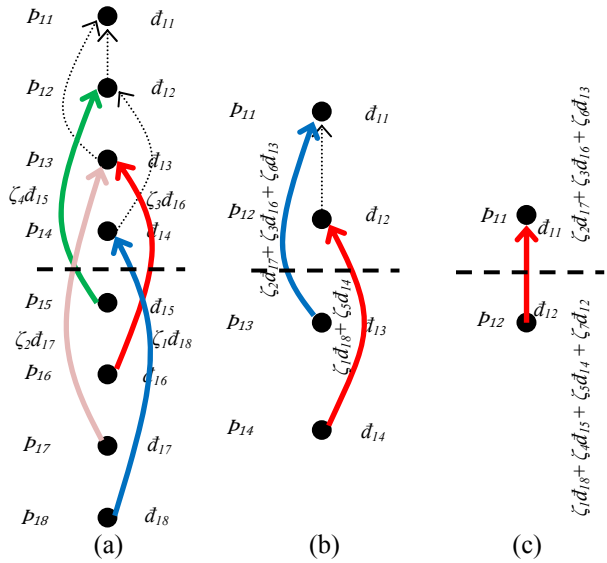

Fig. 13. (a) Iteration first of step 3; data from processers $P_{15}$, $P_{16}$, $P_{17}$ and $P_{18}$ is sent to processors to $P_{14}$, $P_{13}$, $P_{13}$ and $P_{12}$ respectively. (b) Iteration second of step 3; data from processers $P_{14}$ and $P_{13}$ is sent to processors to $P_{12}$ and $P_{11}$ respectively. (c) Iteration third of step 3; data from processers $P_{12}$ is sent to processors $P_{11}$.

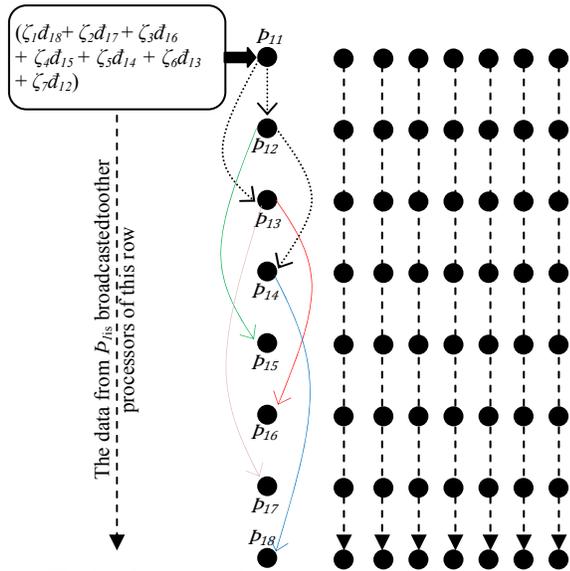

Fig. 14. The data from each column root processor is broadcasted to other processors of respective column in each block.

*Step 7:* Call one–to–all algorithm [19] in the block to transfer the INFO of other blocks (of respective rows) in $n^3 \log n$ time. At the end of this step, complete blocks of each row have INFO of all the processors in that row. Time complexity of step 7: $n^3 \log n$.

*Step 8:* This step performs the interblock communication using horizontal link transfer that transfers the INFO (of all the blocks of respective column) to the root processors (of respective block) with P_ID ($i=n$), and this requires one communication step. Time complexity of step 8: *1CS*.

*Step 9:* Using step 1 transfer of INFO of all the processors with P_ID ($i=n$). Time complexity of step 9: $n^4 \log n$.

*Step 10:* Call AAB algorithm in the block to transfer the INFO of other blocks that column in the block with $n^4 \log n$ time complexity. Time complexity of step 10: $n^4 \log n$.

At the end of this step all, the processors of each block have the INFO of all processors of other blocks.

## 5. Results and Simulations

The implementation of linear coding using AAB on MMT enables the sharing of data between multiple processors, at a time unit, more convenient and easy. As the algorithm becomes more complex, the involvement of processors also increases. For parallel architectures, important issue is to make these architectures more processor utilitarian, otherwise the processors in these architectures are idle, and all are not in use at every step of algorithms. Also, the involvement of coefficients used to broadcast data is high, compared to coefficients involvement after implementation of LCM-PA with AAB on MMT. This makes the algorithm less complex as fewer amounts of coefficients are used for broadcasting data using linear coding. While broadcasting the data in AAB, the time involved to communicate and deliver/receive data from different processors is more. The fall of time complexity at different number of processors shows that the architecture is possible with a set of processors having a combination which makes the algorithm to be implemented with positive results.

The algorithm starts with the execution of each step in the order defined (as step 1... step 10), as the execution of each step starts the involvement of each processors also increases to broadcast data. In parallel processing the algorithm starts with active processor and involves other processors as it progresses [22]. Figure 15 illustrate the involvement of processors with average percentage of iteration in each step.

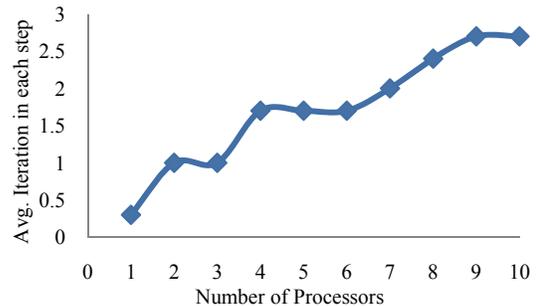

Fig. 15. Involvement of processors at different steps of algorithm.

Based on the above result in figure 15, as the iterations increases the involvement of processors also increases. The algorithm with LCM-PA approach, utilizes the maximum number of processors compared to without LCM-PA approach. So the utilization of processors in parallel architectures is also increases while using linear coding.

## 6. Conclusion and Future Work

We have presented a LCM-PA, model of linear coding, on parallel architecture with efficient implementation of our approach on AAB algorithm on MMT, with comparative time complexity after implementation with LCM-PA. Our model is network independent and can be implemented on any parallel architecture with assumptions to be common as we have used in section second. Future work includes extensions to this approach and analyzing the complexity aspects by implementing with other parallel algorithms (e.g. Multi-Sort [23]). In addition, to make the extension of this approach with LCM-PA model it is needed to be implemented with other parallel algorithms to make vision of research more clear.

**Appendix**

Here we provide the proof of all theorems, definitions and terms used with main text. The definitions used in this paper are defined by other authors but for readers convenience they are elaborated with proof in this section.

**Definition 1** (*Horizontal intrablock links*). The processors in row $i$ of each block $B(\alpha, \beta)$ are connected to form a binary tree rooted at $(\alpha, \beta, i, 1), 1 \leq i \leq n$. That is, for $j = 1$ to $\lfloor n/2 \rfloor$ processor $P(\alpha, \beta, i, j)$ is directly connected to the processors $P(\alpha, \beta, i, 2j)$ and $P(\alpha, \beta, i, 2j + 1)$, whenever they exist.

**Proof.** If this network is used for $N$ number of processors than this type of link exists. Suppose $N = 4$, then total number of processors in the network are $N^4 = 256$ processors, which are divide in four rows and four columns and each row and column consists of four block, and each block consists of four rows and four columns. Now according to definition 1, the processors of block $B(1,1)$ are connected in order:
$P(1,1,1,1) \rightarrow P(1,1,1,2)$ and $P(1,1,1,3)$;
$P(1,1,1,2) \rightarrow P(1,1,1,4)$; //as $P(1,1,1,5)$ does not exist.
$P(1,1,2,1) \rightarrow P(1,1,2,2)$ and $P(1,1,2,3)$;
$P(1,1,2,2) \rightarrow P(1,1,2,4)$;
$P(1,1,3,1) \rightarrow P(1,1,3,2)$ and $P(1,1,3,3)$;
$P(1,1,3,2) \rightarrow P(1,1,3,4)$;
$P(1,1,4,1) \rightarrow P(1,1,4,2)$ and $P(1,1,4,3)$;
$P(1,1,4,2) \rightarrow P(1,1,4,4)$;
As the values of $i$ and $j$ changes the number of connecting horizontal link also varies.°

**Definition 2** (*Vertical intrablock links*). The processors in column $j$ of each block $B(\alpha, \beta)$ are also used to form a binary tree rooted at $(\alpha, \beta, 1, j), 1 \leq j \leq n$. That is, for $i = 1$ to $\lfloor n/2 \rfloor$ processor $P(\alpha, \beta, i, j)$ is directly connected to the processors $P(\alpha, \beta, 2i, j)$ and $P(\alpha, \beta, 2i + 1, j)$, whenever they exist.

**Proof.** If this network is used for $N$ number of processors than this type of link exists. Suppose $N = 4$, then total number of processors in the network are $N^4 = 256$ processors, which are divide in four rows and four columns and each row and column consists of four block, and each block consists of four rows and four columns. Now according to definition 2, the processors of block $B(1,1)$ are connected in order:
$P(1,1,1,1) \rightarrow P(1,1,2,1)$ and $P(1,1,3,1)$;
$P(1,1,2,1) \rightarrow P(1,1,4,1)$; //as $P(1,1,5,1)$ does not exist.
$P(1,1,1,2) \rightarrow P(1,1,2,2)$ and $P(1,1,3,2)$;
$P(1,1,2,2) \rightarrow P(1,1,4,2)$;
$P(1,1,1,3) \rightarrow P(1,1,2,3)$ and $P(1,1,3,3)$;
$P(1,1,2,3) \rightarrow P(1,1,4,3)$;
$P(1,1,1,4) \rightarrow P(1,1,2,4)$ and $P(1,1,3,4)$;
$P(1,1,2,4) \rightarrow P(1,1,4,4)$;
As the values of $i$ and $j$ changes the number of connecting horizontal link also varies.°

**Definition 3** (*Horizontal interblock links*). $\forall \alpha, 1 \leq \alpha \leq n$, the processor $P(\alpha, \beta, i, 1)$ is directly connected to the processor $P(\alpha, i, \beta, n), 1 \leq i, \beta \leq n$. It can be noted that for $\beta = i$, these links connect two processors within the same block.

**Proof.** These are the links between the boundary or corner processors of different blocks. If this network is used for $N$ number of processors than this type of link exists. Suppose $N = 4$, according to definition 3, the processors for $1 \leq \alpha \leq n$ are connected in order:
$P(1,1,1,1) \rightarrow P(1,1,1,4); P(1,1,2,1) \rightarrow P(1,2,1,4);$
$P(1,1,3,1) \rightarrow P(1,3,1,4); P(1,1,4,1) \rightarrow P(1,4,1,4);$
$P(1,2,1,1) \rightarrow P(1,1,2,4); P(1,3,1,1) \rightarrow P(1,1,3,4);$
$P(1,4,1,1) \rightarrow P(1,1,4,4); P(2,1,2,1) \rightarrow P(2,2,1,4);$
As the values of $\beta$, and $i$ changes the number of connecting horizontal links also varies.°

**Definition 4** (*Vertical interblock links*). $\forall \beta, 1 \leq \beta \leq n$, the processor $P(\alpha, \beta, 1, j)$ is directly connected to the processor $P(j, \beta, n, \alpha), 1 \leq j, \alpha \leq n$. It can be noted that for $\alpha = j$, these links connect two processors within the same block.

**Proof.** These are the links between the boundary or corner processors of different blocks. If this network is used for $N$ number of processors than this type of link exists. Suppose $N = 4$, according to definition 3, the processors for $1 \leq \beta \leq n$ are connected in order:
$P(1,1,1,1) \rightarrow P(1,1,4,1); P(1,1,1,2) \rightarrow P(2,1,4,1);$

$P(1,1,1,3) \rightarrow P(3,1,4,1); P(1,1,1,4) \rightarrow P(4,1,4,1);$
$P(2,1,1,1) \rightarrow P(1,1,4,2); P(3,1,1,1) \rightarrow P(1,1,4,3);$
$P(4,1,1,1) \rightarrow P(1,1,4,4);$

As the values of $\alpha$ and $i$ changes the number of connecting vertical links also varies.°

**Definition 5** (*Directed Graph*). A parallel network in any of the phase of communication is said to be directed based on the flow of data with respect to the algorithm.

**Proof.** A parallel network is said to be directed, when the flow of data is decided based on some parameters. As MMT network is bidirectional, in some part of communication it is using a specific orientation for communication while in some parts it may be reverse, based on the algorithm used to decide the communication. As an example consider algorithm 1 in which the communication is performed from greater processor index to lesser processor index, so the direction is different from algorithm 2 in which the root processor transfers data to other processors of respective rows.°

**Nitin Rakesh** is Sr. Lecturer in the Department of Computer Science and Engineering & Information Technology, Jaypee University of Information Technology (JUIT), Waknaghat, Solan–173215, Himachal Pradesh, India. He was born on October 30, 1982, in Agra, India. In 2004, he received the Bachelor's Degree in Information Technology and Master's Degree in Computer Science and Engineering from Jaypee University of Information Technology, Noida, India in year 2007. Currently he is pursuing his doctorate in Computer Science and Engineering and his topic of research is parallel and distributed systems. He is a member of IEEE, IAENG and is actively involved in research publication. His research interest includes Interconnection Networks & Architecture, Fault–tolerance & Reliability, Networks–on–Chip, Systems–on–Chip, and Networks–in–Packages, Network Algorithms, Parallel Algorithms, Fraud Detection. Currently he is working on Efficient Parallel Algorithms for advanced parallel architectures.

**Dr. Vipin Tyagi** *is* Associate Prof. in Department of Computer Science and Engineering at Jaypee University of Information Technology, Waknaghat, India. He has about 20 years of teaching and research experience. He is an active member of Indian Science Congress Association and President of Engineering Sciences Section of the Association. He is a Life Fellow of the Institution of Electronics and Telecommunication Engineers and a senior life member of Computer Society of India. He is member of Academic-Research and Consultancy committee of CSI. He is elected as Executive Committee member of Bhopal Chapter of CSI and M.P. and CG chapter of IETE. He is a Fellow of Institution of Electronics and Telecommunication Engineers, life member of CSI, Indian Remote Sensing Society, CSTA, ISCA and IEEE, IAENG. He has published more than 50 papers in various journals, advanced research series and has attended several national and international conferences in India and abroad. He is Principal Investigator of research projects funded by DRDO, MP Council of Science and Technology and CSI. He has been a member of Board of Studies, Examiner Member of senate of many Universities. His research interests include Parallel Computing, Image Processing and Digital Forensics.